\documentclass[12pt, preprint]{aastex}

\usepackage{epsf,color,amsmath}

\newcommand{\sfrac}[2]{\mathchoice%
  {\kern0em\raise.5ex\hbox{\the\scriptfont0 #1}\kern-.15em/
    \kern-.15em\lower.25ex\hbox{\the\scriptfont0 #2}}
  {\kern0em\raise.5ex\hbox{\the\scriptfont0 #1}\kern-.15em/
    \kern-.15em\lower.25ex\hbox{\the\scriptfont0 #2}}
  {\kern0em\raise.5ex\hbox{\the\scriptscriptfont0 #1}\kern-.2em/
    \kern-.15em\lower.25ex\hbox{\the\scriptscriptfont0 #2}} {#1\!/#2}}

\newcommand{\myhalf}{\sfrac{1}{2}}

\newcommand{\eb}{{\bf{e}}}
\newcommand{\Ub}{{\bf{U}}}
\newcommand{\xb}{{\bf{x}}}
\newcommand{\kb}{{\bf{k}}}
\newcommand{\Vb}{{\bf{V}}_n}
\newcommand{\Vbhat}{{\bf{\widehat{V}}}_n}

\usepackage{multirow}

\newcommand{\Ubt}{\widetilde{\Ub}}

\newcommand{\isot}[2]{$^{#2}\mathrm{#1}$}
\newcommand{\isotm}[2]{{}^{#2}\mathrm{#1}}

\newcommand{\maestro}{{\sf MAESTRO}}
\newcommand{\castro}{{\sf Castro}}
\newcommand{\boxlib}{{\sf BoxLib}}

\newcommand{\gcc}{\mathrm{g~cm^{-3} }}

\newcommand{\uadv}{\widetilde{\mathbf{U}}^{\mathrm{ADV}}}
\def\uadvstar {\mathbf{U}^{\mathrm{ADV},\star}}
\def\uadvpred {\mathbf{U}^{\mathrm{ADV},\mathrm{pred}}}

\def\ib    {{\bf i}}

\newcommand{\pa}{\ensuremath{\left(\mathrm{p},\alpha\right)}}

\begin{document}
\title{Comparisons of Two- and Three-Dimensional Convection in Type I X-ray Bursts}

\shorttitle{Comparisons of Two- and Three-dimensional Convection in XRBs}
\shortauthors{Zingale et al.}

\author{M.~Zingale\altaffilmark{1},
        C.~M.~Malone\altaffilmark{2,4},
        A.~Nonaka\altaffilmark{3},
        A.~S.~Almgren\altaffilmark{3},
        J.~B.~Bell\altaffilmark{3}}
\email{michael.zingale@stonybrook.edu}

\altaffiltext{1}{Dept.\ of Physics \& Astronomy,
                 Stony Brook University,
		 Stony Brook, NY 11794--3800}

\altaffiltext{2}{CCS-2 \& XCP-1,
                 Los Alamos National Laboratory, 
                 Los Alamos, NM 87545}

\altaffiltext{3}{Center for Computational Sciences and Engineering,
                 Lawrence Berkeley National Laboratory,
                 Berkeley, CA 94720}

\altaffiltext{4}{Nicholas C.~Metropolis Fellow}

\begin{abstract}
We perform the first detailed three-dimensional simulation of low Mach
number convection preceding runaway thermonuclear ignition in a mixed H/He
X-ray burst.  Our simulations include a moderate-sized, approximate
network that captures hydrogen and helium burning up through
rp-process breakout.  We look at the difference between two-
and three-dimensional convective fields, including the details of the
turbulent convection.
\end{abstract}


\section{Introduction}\label{Sec:Introduction}

X-ray bursts (XRBs) are the thermonuclear runaway in a H/He layer on
the surface of a neutron star.  These transient events can be used to
probe the structure of neutron stars and the equation of state of
dense material~\citep{steiner:2010,ozel:2010}.  Furthermore, they are
also the sites of rp-process nucleosynthesis~\citep{rpprocess}.  For
these reasons, understanding the dynamics of the explosion has seen
substantial research interest in the past years.

One-dimensional studies
\citep{taam1980,taamwoosleylamb1996,woosley-xrb} can reproduce the
observed energies, durations, and recurrence timescales for XRBs, but
use a parameterized model for convection, namely mixing length theory
(which likely does not describe turbulent convection accurately
e.g.~\citealt{arnett:2015}).  An open question is whether a
fully-turbulent convective velocity field can modify the
nucleosynthesis.  Additionally, the convection may dredge up heavy
element ash to the photosphere~\citep{intZand:2010,bhattacharyya:2010}
thereby altering the opacity of the atmosphere, which affects the
inference of neutron star mass and radius from photospheric radius
expansion (PRE) bursts.  These are inherently three-dimensional
problems.

Previously, we performed two-dimensional simulations, focusing first
on pure He bursts \citep{xrb}, and then later on mixed H/He bursts
\citep{xrb2}.  The latter study used an approximate network to capture
the hot-CNO, triple-$\alpha$, and initial rp-process breakout burning.
There we found that we needed a spatial resolution of about 6~cm
zone$^{-1}$ in order to accurately model the burning; for comparison,
the extremely temperature-sensitive burning of the pure He models of
\cite{xrb} required 0.5~cm zone$^{-1}$ resolution.  In this paper, we
extend our studies by performing the first three-dimensional model of
convective burning in a H/He XRB, using the reaction network from
\citet{xrb2}.  This initial study compares to our two-dimensional
results, and discusses the computational requirements for a more
extensive study.

\section{Numerical Method}\label{Sec:Numerical Method}

We use the publicly-available\footnote{\maestro\ can be obtained from
  {\tt http://github.com/BoxLib-Codes/MAESTRO/}}
\maestro\ code~\citep{MAESTRO:Multilevel}, which solves the equations
of low Mach number hydrodynamics by reformulating the reactive Euler
equations to filter soundwaves while retaining compressibility effects
due to stratification and local heat release.  By filtering
dynamically unimportant soundwaves, \maestro\ enables efficient
simulation of slow convective flows, such as those in
XRBs~\citep{xrb,xrb2}, various progenitors of Type Ia
supernovae~\citep{wdconvect,wdturb,subchandra}, and in the cores of
massive stars~\citep{ms_cc}.  Also important for simulations like
these is that the low Mach number formulation analytically enforces
hydrostatic equilibrium of the base state, allowing us to maintain a
hydrostatic atmosphere in the simulation code without the development
of large spurious velocities (see, e.g., \citealt{ppm-hse}).

All of the \maestro\ options and microphysics used in our
two-dimensional study of XRBs in \citet{xrb2} are retained for this
study.  In particular, we use the new energy formulation variant of
\maestro, based on the ideas in \citet{kleinpauluis} and
\citet{vasil:2013}, which improves energy conservation and our
treatment of gravity waves.  We use the Helmholtz equation of
state (EOS) from \citet{timmes_swesty:2000}, which includes an ideal
gas of nuclei, a photon gas, and an electron/positron gas with
arbitrary degeneracy and relativistic parameters, and Coulomb
corrections.  \maestro\ is under continuous development, and
we've improved the advection portion of the code since the
construction of the interface states was last discussed in
\citet{ABNZ:III}.  We take the opportunity to document those changes in
Appendix~\ref{Sec:Godunov Integration Details}.

\begin{figure}
\centering
\plotone{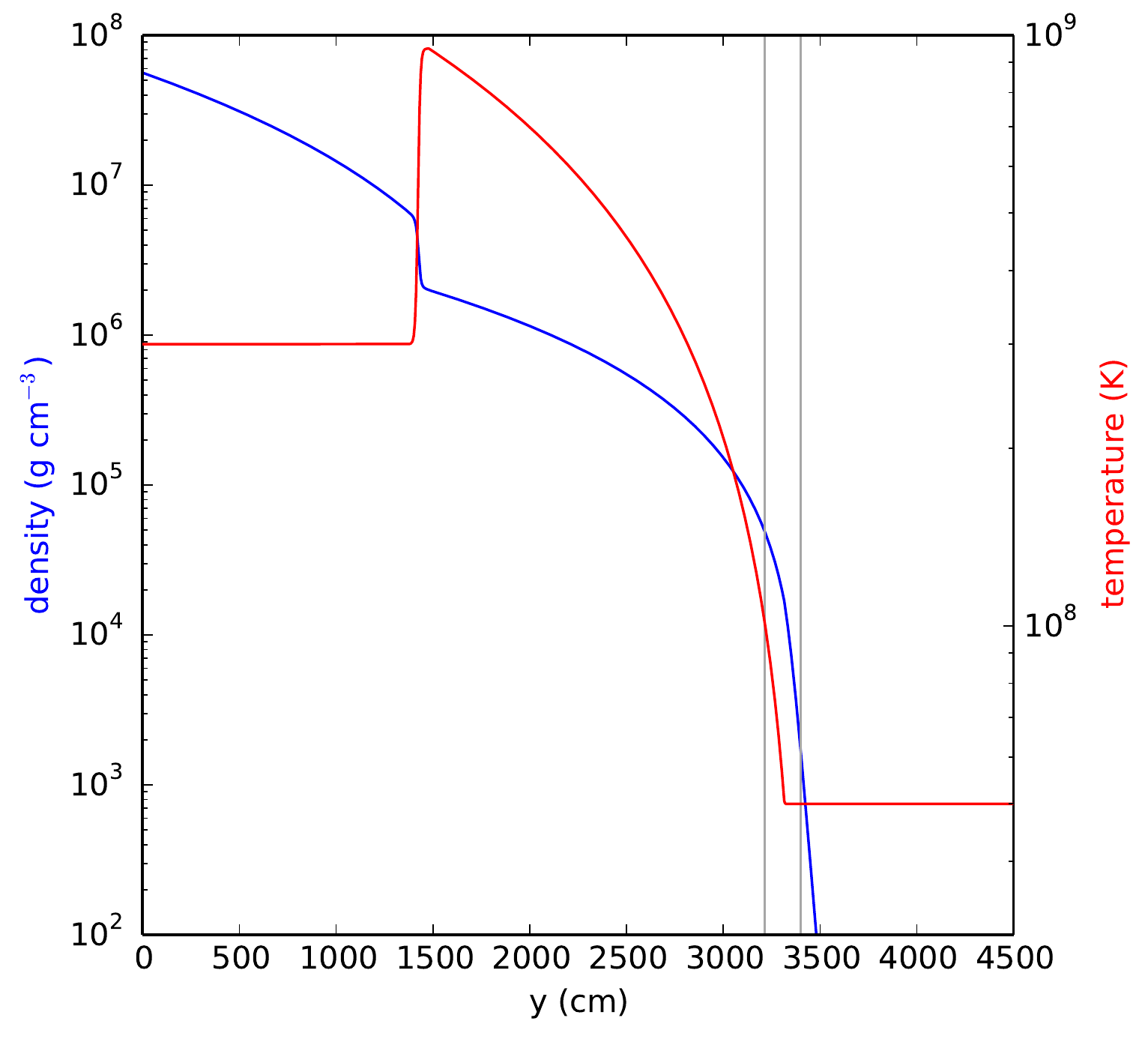}
\caption{\label{fig:initial_model} Initial density and temperature profile.  The
  vertical lines represent the sponge start (leftmost line) and the anelastic cutoff
  for the three-dimensional runs.}
\end{figure}

We use the same parametrized initial model as in our two-dimensional
study.  Briefly, the model consists of a $M = 1.4~M_\odot$, $R =
10$~km neutron star, of which we model the outer
$\sim1.4\times10^3$~cm as an isothermal ($T = 3\times10^8$~K), pure
\isot{Ni}{56} gas.  On top of the neutron star is a warm accreted
layer of mainly H/He fuel that is slightly metal-rich compared to
solar, with CNO metals tied up in \isot{O}{14} and \isot{O}{15} in a
ratio comparable to their respective $\beta$-decay lifetimes.  A
smooth transition is applied between the density ($\rho =
2\times10^6$~g cm$^{-3}$) and temperature ($T=9.5\times10^8$~K) at the
base of the accreted layer and the surface of the neutron star.  The
accreted layer is given an isentropic profile, making it convectively
unstable, and the temperature decreases until a cutoff temperature is
reached.  The original extent of the convective region is $\lesssim
2\times10^3$~cm.
Figure~\ref{fig:initial_model} shows the density and temperature
profile, along with the values of the cutoff densities that are part
of the \maestro\ algorithm.  For the three-dimensional simulations
present here, the anelastic and base cutoff densities have been
increased slightly to $2\times 10^3~\gcc$ to better quench the
dynamics above the atmosphere.
The two-dimensional simulations used the same parameters as in
\citet{xrb2}.  The reader is referred to the Appendix of \citet{xrb2}
for more details of our model construction procedure.  Finally we note
that all of the problem setup files, inputs, and initial models for
the runs presented here have been copied into the main \maestro\ code
repository in {\tt Exec/SCIENCE/xrb\_mixed/}, allowing anyone to rerun
these simulations.

In this paper, we perform two three-dimensional simulations to assess
the dynamics of the convective flow.  We model the XRB using a
plane-parallel geometry.  Our wide simulation uses a uniform grid of
$512\times 512\times 768$ and our narrow simulation uses a grid of
$256\times 256\times 768$ zones, both with 6~cm zone$^{-1}$ spatial
resolution---the same resolution used in our two-dimensional study.
As the simulation evolves, the one-dimensional hydrostatic base state
that \maestro\ carries is allowed to expand due to the heating,
following the procedure described in \citet{ABRZ:II}.

\subsection{Correction to the Network}\label{SSec:network_corrections}
Our reaction network contains 10 species, approximating hot CNO,
triple-$\alpha$, and rp-breakout burning up through \isot{Ni}{56},
using the ideas from \citet{wallacewoosley:1981}, but with modern
reaction rates from {\sf ReacLib}~\citep{ReacLib} where available (see
the discussion in \citealt{xrb2} for more details).  This is the same
network used in \citet{xrb2}, but with one important change.
The convective flow field in \citet{xrb2} showed signs of splitting
into two distinct convective regions (e.g. Figure 7 of that paper).
The split occured at a location of a secondary peak in energy
generation, which grew with time (Figure 9 in that paper).  We
attributed this extra energy to the branching ratio, $\lambda_1$, of
$\beta$-decay versus $\alpha$-capture on \isot{Ne}{18} as a breakout
mechanism from the Hot CNO cycle.  The precise location of the
secondary peak in energy production was where the branching ratio
favored the $\beta$-decay to \isot{F}{18} (see Figure 10 of
\citealt{xrb2}), followed by $\isotm{F}{18}\pa\isotm{O}{15}$; the approximate 
network converts \isot{F}{17}$+2$p directly to \isot{O}{15}$+\alpha$
at a rate governed by the rate of p-capture on \isot{F}{17} and
$\lambda_1$.

This coincidence of peak energy generation and $\lambda_1$ transition
was a red herring: the energy generation from the
\isot{F}{17}$\left(2\mathrm{p},\alpha\right)$\isot{O}{15} chain was
insufficient to reproduce the production rate we witnessed.  We know
now that we erroneously had an additional term in the reaction network
--- based on legacy code --- that attempted to model p-capture on
\isot{Ni}{56} to heavier elements.  In particular, there was a kludge
of a term involving
$\isotm{Ni}{56} + 56\mathrm{p} \rightarrow 2\, \isotm{Ni}{56}$ to mimic the
energy release of heavier element production, which should not
have been included in the network.  This ``reaction'' occured exactly
at the secondary peak in energy generation and depletion of H, and its
rate was sufficient to reproduce the energy production and its
increase with time.  We have since removed this feature of our
network.  All calculations in this paper, including the 2-d comparisons,
use the corrected network,
which is available in the \maestro\ distribution in {\tt
  Microphysics/networks/rprox/}.

\section{Results}\label{Sec:Results}

\begin{figure}
\centering
\plotone{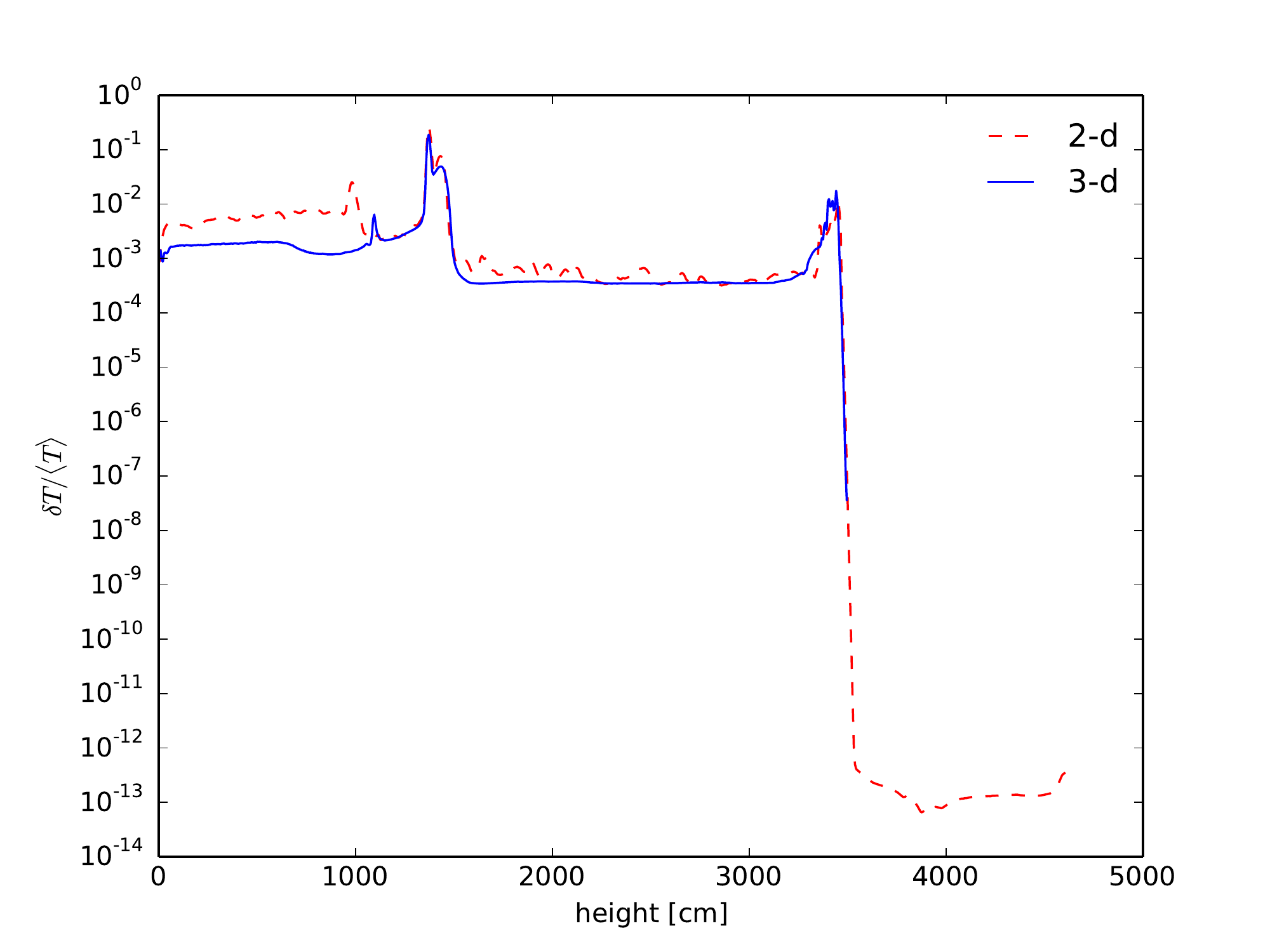}
\caption{\label{fig:dT} Variance of $T$ normalized by its average at a
  given height for both our two-dimensional and three-dimensional
  simulations at $t=0.02$~s.  Within the convective region, $1600
  \lesssim \textrm{height} \lesssim 3200$, the temperature fluctuations
  between the two simulations are quite similar.  We also note that at
  the edges of the convective region, due to overshoot/undershoot,
  there are local spikes in the average temperature fluctuations.
  Below the convective region, the two-dimensional simulation shows
  temperature fluctuations that are about four times larger than in
  the three-dimensional counterpart.  This is likely due to the larger
  amount of convective undershoot present in the two-dimensional
  simulation compared to the three-dimensional simulation.
  Note that the variation for the 3-d simulation is 0 above 3500 cm,
  and as a result the line is not plotted on the log scale.}
\end{figure}

In order to understand how dimensionality affects our results, we
compare to updated two-dimensional calculations based on \citet{xrb2}.
In particular, we use a 6~cm resolution $1024\times 768$ zone
calculation.  Figure~\ref{fig:dT} shows the standard deviation of
temperature (compared to other zones at the same height) as a function
of height for the two- and three-dimensional runs, both at $t =
0.02$~s.  The overall trend is the same for the two calculations, with
the magnitude of the temperature fluctuations in the convective region
($\sim$ 1400 cm to 3550 cm) $\delta T/\langle T\rangle \sim 10^{-3}$
to $10^{-4}$.

\begin{figure}
\centering
\plottwo{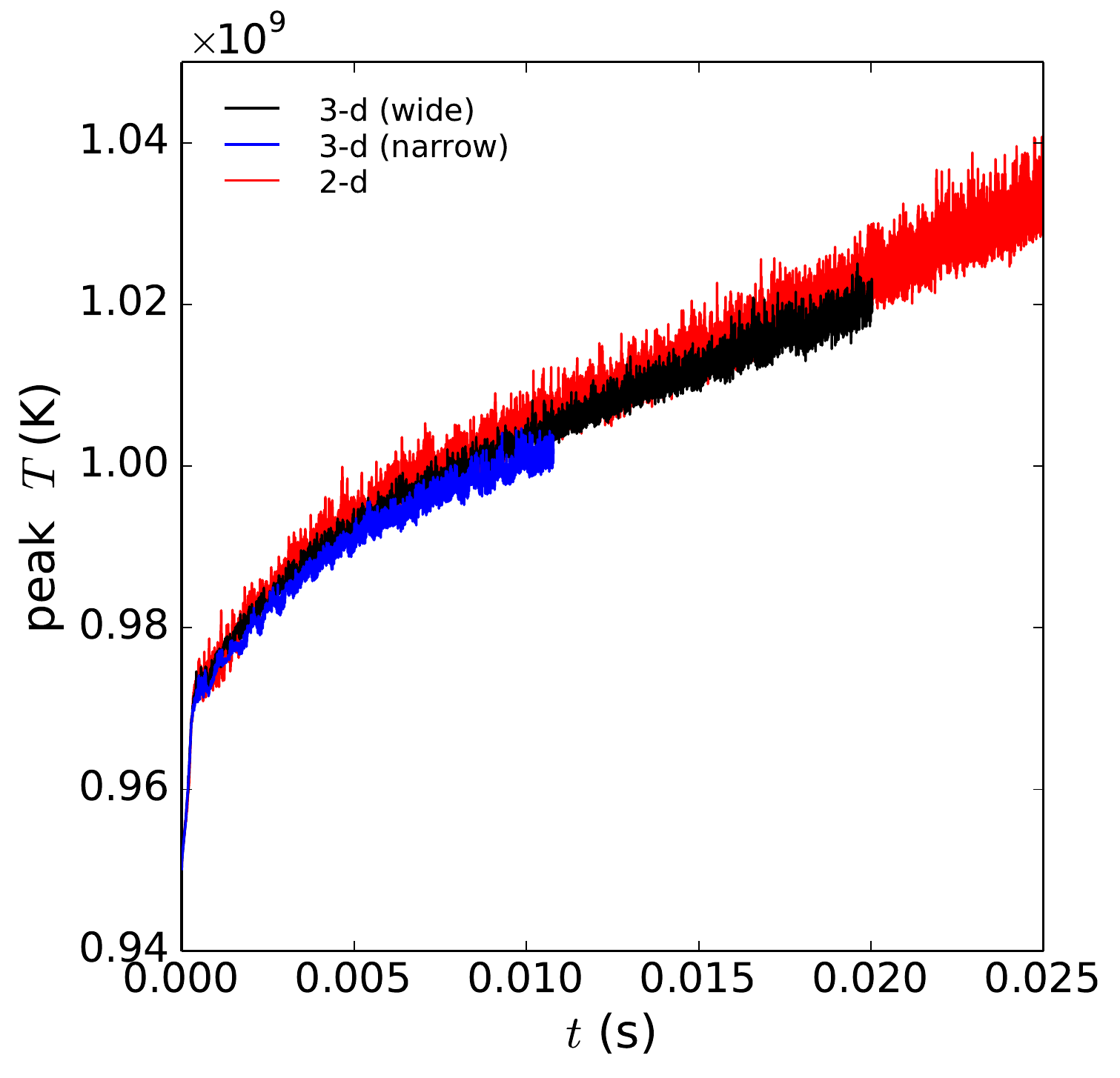}{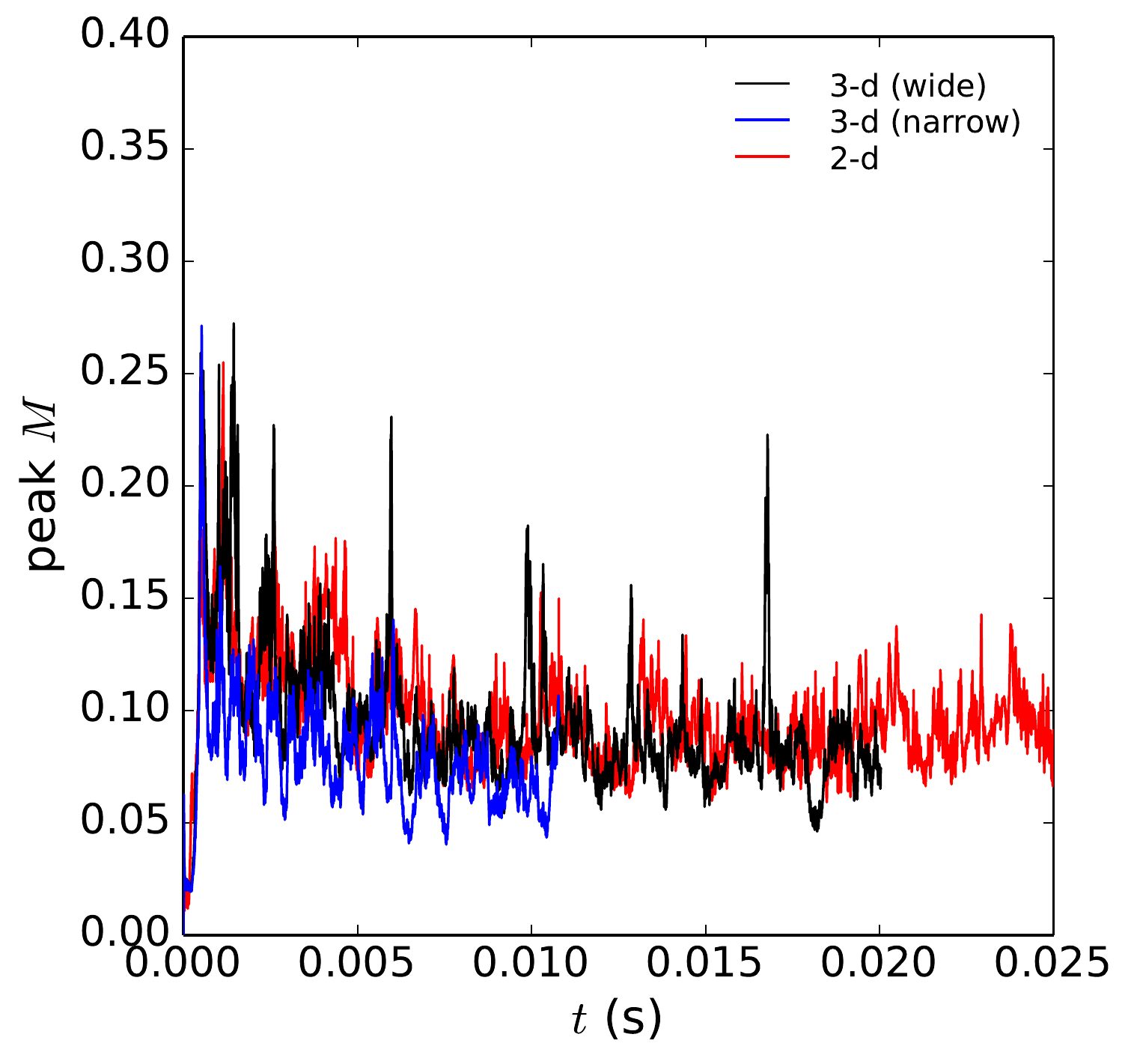}
\caption{\label{fig:temp} Comparison of the peak temperature vs.\ time
  between the two- and three-dimensional simulations (left) and the
  peak Mach number vs.\ time (right).  All simulations agree quite
  well in this context, however the three-dimensional simulation has
  more spikes to large Mach number at late times.  All simulations
  experience an initial short-duration transient spike in Mach number
  as the system creates a convective flow field able to carry away the
  energy generated from nuclear reactions.}
\end{figure}

Figure~\ref{fig:temp} shows the peak temperature and peak Mach number
as a function of time for the runs.  We see that they closely track
one another, but that in the wide three-dimensional simulation there
more sporatic spikes to moderate Mach number throughout the simulation.
At the start
of the calculation, there is always a period of transient behavior as
the heating needs to set up a consistent convective velocity field,
but the flow quickly settles down.  For both simulations, the average
Mach number after the transient is about 0.1; in the longer-duration
two-dimensional case, the Mach number asymptotes to 0.1.
The temperature plots all track one another well.
We did not
run the three-dimensional calculation as long as the two-dimensional
calculation, to conserve computational resources.

It is interesting to note how the peak Mach number translates into a
timestep improvement compared to a fully compressible code.  For this
problem, the sound speed in the atmosphere is greater the deeper one
goes into the atmosphere, but the Mach number is highest at the top of
the atmosphere.  As a result, the timestep increase will actually be
better than the naive $1/M$ one would expect if the domain were
uniform.  A further complication is that, in the compressible code,
when we reach the low density material at the top of the atmosphere
that buffers us from the boundary, it is radiation pressure that
dominates here, articially increasing the hydrodynamic soundspeed.
This is a common limitation that arises from using an EOS that
includes radiation instead of modeling the radiation field itself.
For comparison, we started the same XRB simulation (in two dimensions)
in the \castro\ code~\citep{castro}, and the average timestep after
$2.76\times 10^{-4}$~s of evolution was $\Delta t_\mathrm{comp} =
2.79\times 10^{-10}$~s.  For the main three-dimensional
\maestro\ calculation, the average timestep over the course of the
entire simulation was $\Delta t_\mathrm{LM} = 1.93\times
10^{-7}$~s---a $\sim 700\times$ improvement. 

\begin{figure}
\centering
\includegraphics[height=7.5in]{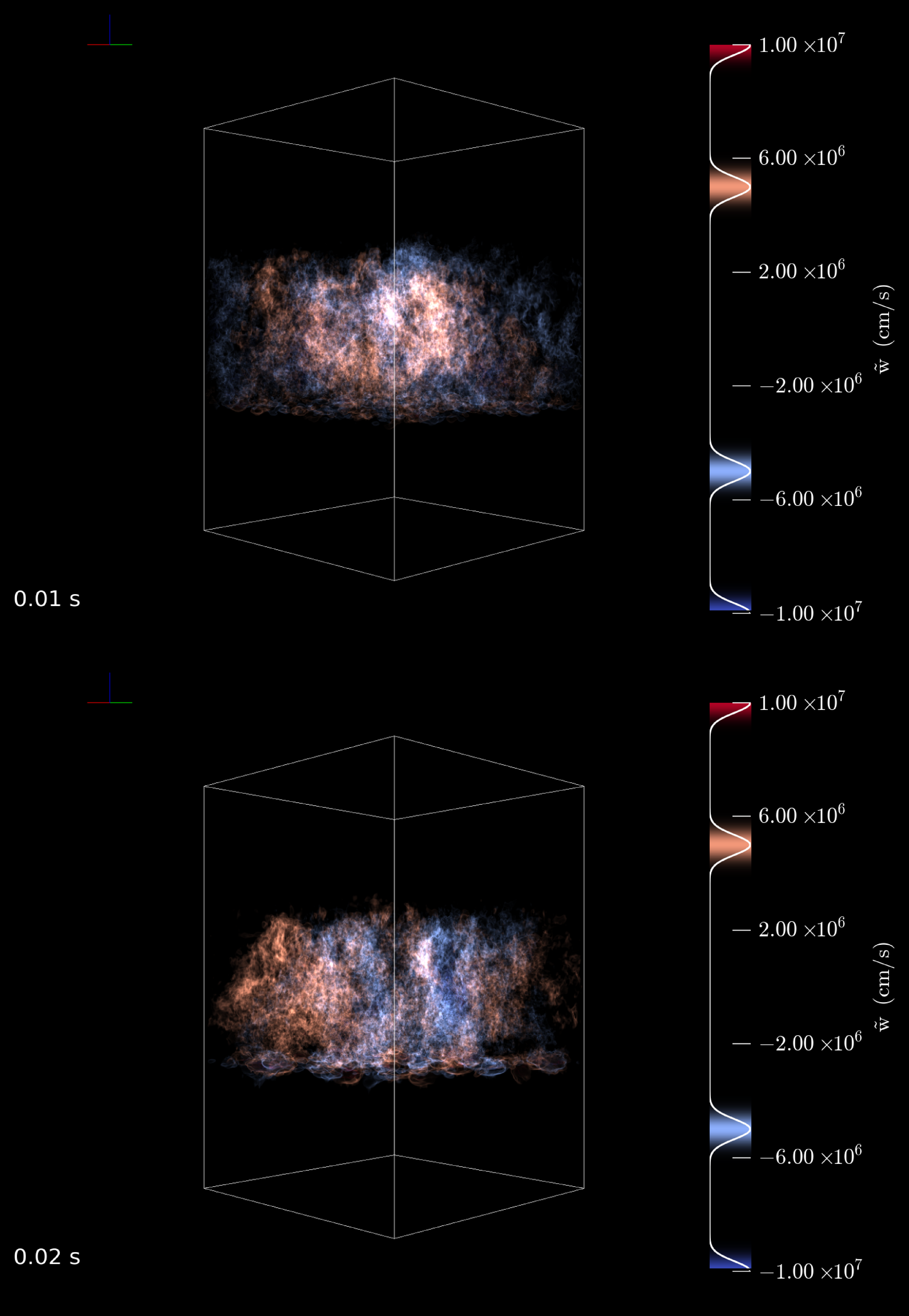}
\caption{\label{fig:vz} Volume renderings of the vertical velocity
  field at $t=$0.01~s (top) and 0.02~s (bottom) for the wide calculation.
  Upward moving fluid is in red and downward moving is
  blue.  }
\end{figure}

The convective velocity structure of the wide three-dimensional
simulation is shown in Figure~\ref{fig:vz}, highlighting the vertical
velocity.  These two images are representative of the flow throughout
the simulation.  We do not see the tight layering that was apparent in
the older two-dimensional simulations (especially for narrower
domains; see Figures 6 and 7, and the discussion in Section 4.2.1 of
\citealt{xrb2}) because of the fix to the reaction network discussed
in Section \ref{SSec:network_corrections}.  To better understand the
difference in the nature of the convective flow, we need to examine
the turbulent structure. 

Turbulence is known to behave differently between two and three
dimensions (see, e.g.~\citealt{ouellette:2012}).  To get a feel for the turbulent nature of the convection
in these simulations, we look at the kinetic energy power spectrum.
Following the discussion regarding turbulence in stratified flows in
\citet{wdturb} and references therein, we calculate a generalized
kinetic energy density spectrum as
\begin{equation}\label{eq:general-spectrum}
E_n(k) = \frac{1}{\Omega} \int_{\mathbf{S}(\kb)} \frac{1}{2}
\Vbhat(\kb) \cdot \Vbhat^\star(\kb) d\mathbf{S},
\end{equation}
where $\Vbhat(\kb)$ is the Fourier transform of $\Vb(\xb) =
\rho^n(\xb) \Ubt(\xb)$ with $n$ specifying the density weighting,
$\mathbf{S}(\kb)$ is the surface defined by $|\kb|=k$, and the
$\star$ denotes complex conjugation.  We note that here we use $\Ubt$,
the local velocity on the grid, instead of explicitly calculating the
turbulent velocity fluctuations from the full velocity field,
including the base state expansion, $\Ub = \Ubt + w_0\eb_r$, because
$\Ubt$ is essentially the velocity perturbations on top of an
otherwise hydrostatic background state~\citep{MAESTRO:Multilevel}.
The volume, $\Omega$, and surface element, $d\mathbf{S}$, are
based on the dimensionality of the problem.  The goal is to find the
proper scaling of the energy density spectrum with wavenumber for both
two- and three-dimensional flow.

The units of $\Vbhat$ are [g$^n$ cm$^{1+D-3n}$ s$^{-1}$], where the
extra power of $D$ on the length scale comes from the integral over
$d\xb$ in the definition of the Fourier transform.  In
Equation~\eqref{eq:general-spectrum}, the integral is done in
$k$-space, such that $d\mathbf{S} \sim d^{D-1}k$ with units
[cm$^{1-D}$], whereas the normalization is in real-space, so that
$\Omega$ has units of [cm$^D$].  Upon integration of
Equation~\eqref{eq:general-spectrum}, the dimensionality, $D$, drops
out of the equation, and the units of the generalized kinetic energy
density spectrum become [g$^{2n}$ cm$^{3-6n}$ s$^{-2}$] for both two-
and three-dimensional configurations.  For turbulent flows that have
density variation (i.e.~compressible or stratified flows), the typical
Kolmogorov energy dissipation rate, $\epsilon(l)$, at a given length
scale $l$ should be weighted by the mass density~\citep[see][for
  example]{Fleck83,Fleck96}: $\epsilon(l) = \rho U^3(l)/l$, which has
units of [g cm$^{-1}$ s$^{-1}$].  The arguments of \citet{wdturb} then
apply to any dimension: the only combination of $\epsilon^\alpha
k^\beta E_n(k)$ that yields a dimensionless quantity is when $\alpha =
-2/3$, $n = 1/3$, and $\beta = 5/3$.  If the physics of two
dimensional and three dimensional turbulence were the same (this is
likely not the case), then the spectrum defined in
Eq.~\eqref{eq:general-spectrum} should scale as $k^{-5/3}$ for
both two- and three-dimensional flows.

In evaluating Eq.~\eqref{eq:general-spectrum}, we create
equally-spaced radial wavenumber bins, $k_i$, ranging from the smallest
physical wavenumber, $1/L$, to the highest meaningful wavenumber,
$1/(2\Delta x)$, where $L$ is the domain width.  The Fourier transform
of the kinetic energy density gives us 
\begin{equation}
\hat{K}(k_x, k_y, k_z) = \frac{1}{2} ( \Vbhat \cdot \Vbhat^\star )
\end{equation}
For each of points in the three-dimensional $\hat{K}$ array, we define
$|k| = \sqrt{k_x^2 + k_y^2 + k_z^2}$ and determine which of the radial
bins, $k_i$, this falls into and add the value of $\hat{K}$ to that
bin's sum.  Done this way, we are integrating up in spherical shells
in $k$-space, using our discrete bins.  The same procedure is done in
two dimensions, but now we are working in the $k_x$-$k_y$ plane, and
are integrating up over annular regions in that plane, again defined
by our discrete bins, $k_i$.  We do not worry about the $1/\Omega$
normalization, since we will normalize each spectrum such that its peak
value is 1.

\begin{figure}
\centering
\plotone{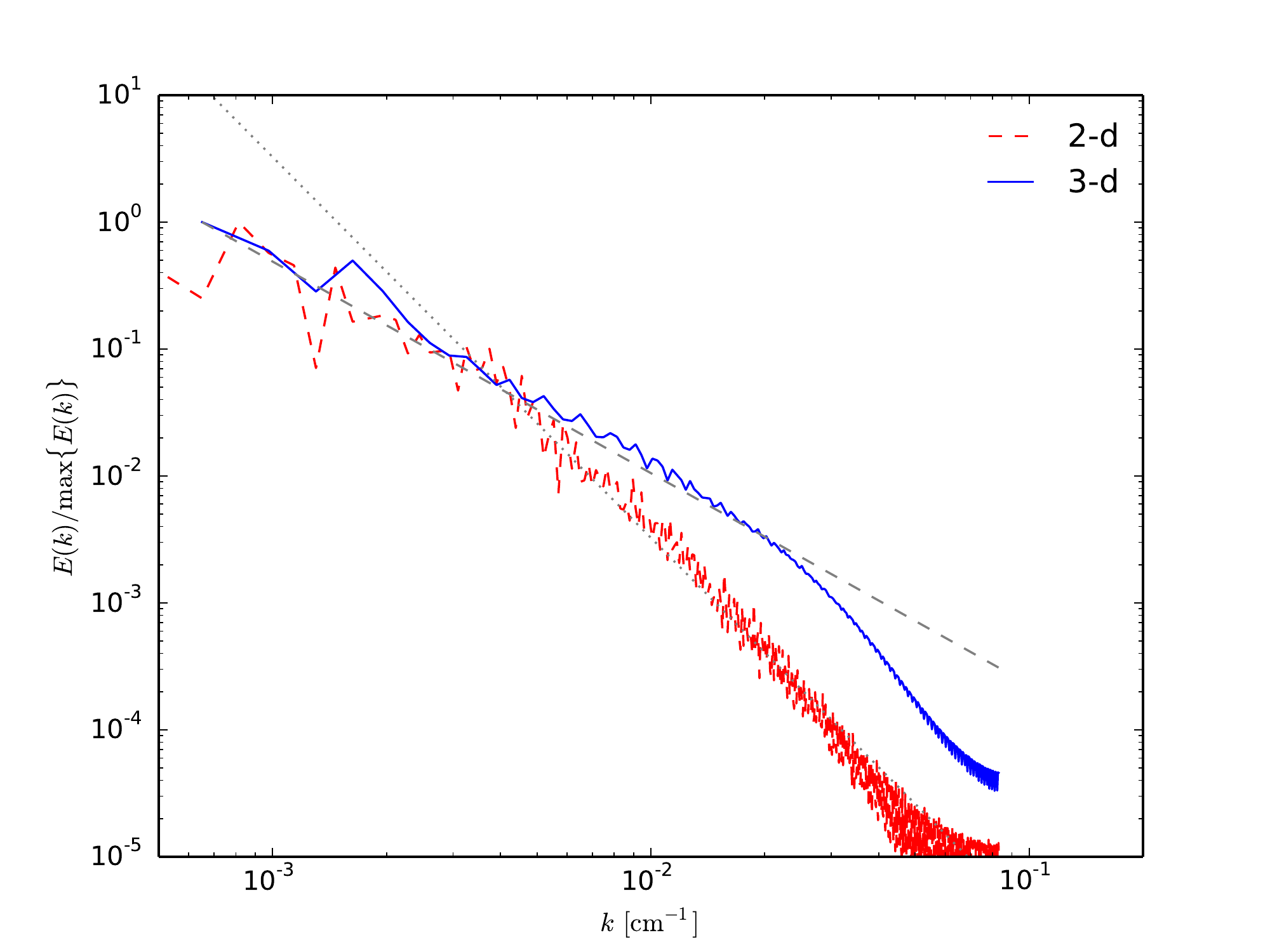}
\caption{\label{fig:turb} Kinetic energy power spectrum for the two-
  and three-dimensional simulations at $t=0.02$~s.  The dashed gray
  line is a $k^{-5/3}$ power-law and the dotted line is a $k^{-3}$
  power-law.  A density weighting of $\rho^{1/3}$ was used for both
  two and three dimensions.  The power is normalized so the two
  spectra have the same peak.  There is about a decade in wavenumber
  where the three-dimensional simulation obeys the standard Kolmogorov
  turbulent cascade.  The two-dimensional simulation displays a
  characteristic change in power-law scaling, having sections that are
  both shallower and steeper than what Kolmogorov predicts for three
  dimensions.}
\end{figure}

Figure~\ref{fig:turb} shows the power spectrum of the two-dimensional
and wide three-dimensional XRB simulations at $t = 0.02$~s.  For this
analysis, we restrict the domain to just the vicinity of the
convective region, including only the vertical range $1300~\mathrm{cm}
< z < 3550~\mathrm{cm}$.  For the three-dimensional case, we see that
we have more than a decade in wavenumber where we achieve a $k^{-5/3}$
power-law scaling, indicative of Kolmogorov turbulence.  We note that
the region we are studying is not periodic in the vertical direction,
but an FFT assumes periodicity, so the discontinuity through the
vertical boundary may affect the behavior at high wavenumbers, perhaps
accounting for the slow fall in the three-dimensional spectrum.  For
the two-dimensional case, the spectrum starts off with a $k^{-5/3}$
scaling, but then becomes steeper at moderate wavenumbers.  Such a
break in the power law scaling for two-dimensional turbulence is
predicted for very idealized turbulence where the steeper part of the
curve has a $k^{-3}$ scaling attributed to a cascade of enstrophy
\citep[e.g.~][]{Kraichnan67,Leith68,Batchelor69}.  Numerical
simulation cannot achieve the idealized conditions (e.g.~infinte
domain and infinite Reynolds number) assumed in the $k^{-3}$
derivation, and sometimes achieve a steeper power law \citep[e.g.~the
  review by][and references therein]{GT06}.  In our two-dimensional
simulation, we see a moderate range in the spectrum after the break
consistent with $k^{-3}$. 

We have also seen such a difference in scaling between two- and
three-dimensional turbulence on smaller scales in reactive
Rayleigh-Taylor simulations~\citep{SciDac2005}, where we saw a
spectrum that appeared to follow the $k^{-11/5}$ scaling predicted by
Bolgiano-Obukhov statistics for a two-dimensional
cascade~\citep{NK:1997}.  In that study, we found that a wide domain,
giving more statistics, was essential to see this scaling.  The
difference in the scaling we observe in the present simulations
suggests that there is a fundamental difference in how the cascade
takes place between the two- and three-dimensional convection in XRBs.

\begin{figure}
  \centering
  \plotone{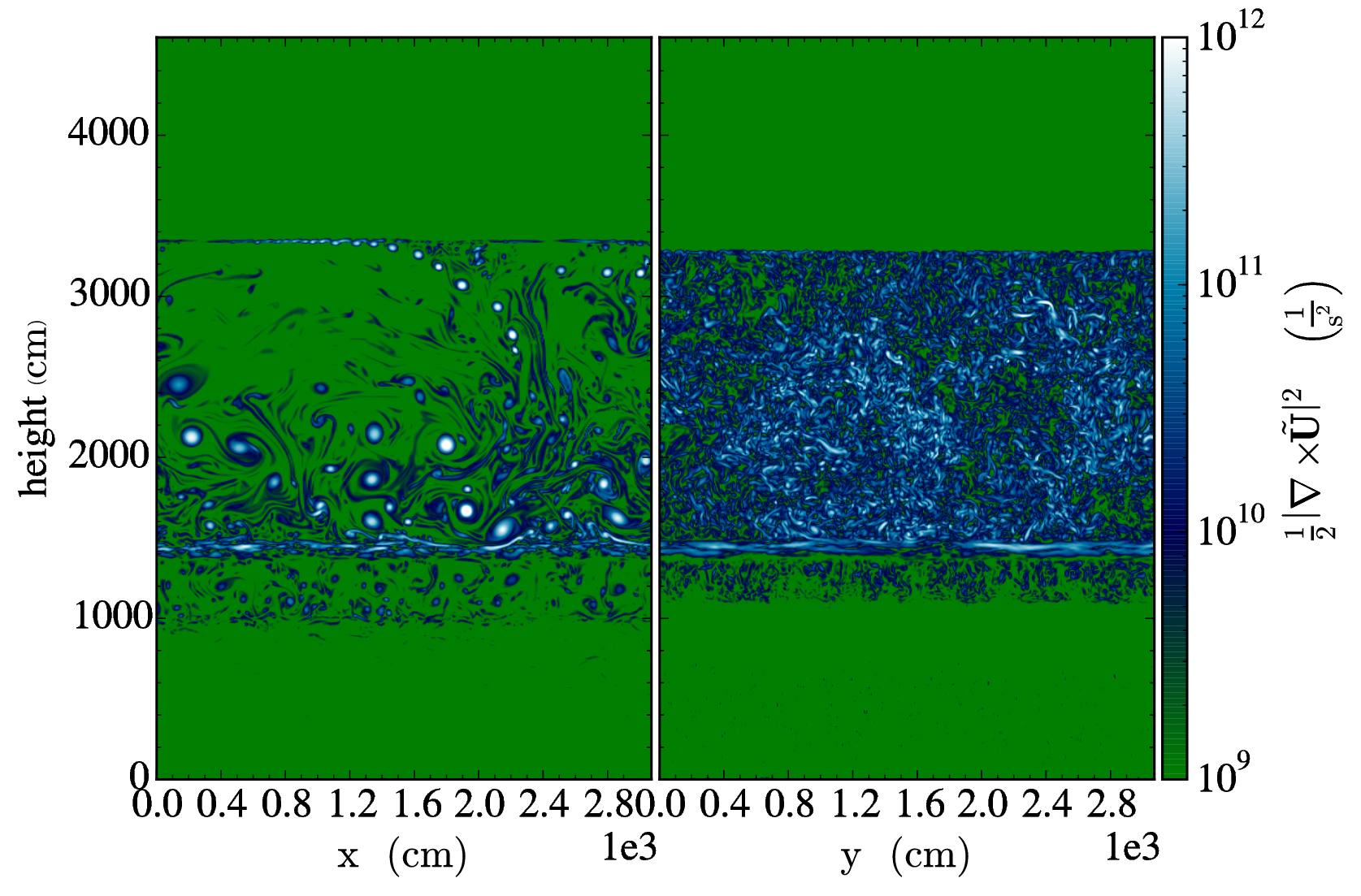}
  \caption{\label{fig:enstrophy} Enstrophy density of the turbulent
    flow in both the two-dimensional (left) and three-dimensional
    (right) simulations at $t=0.02$~s.  The plot for the
    three-dimensional simulation is a slice through the center of the
    domain; the plot for the two-dimensional simulation only shows
    half of the wide domain to keep the spatial scale the same in both
    plots.  There are clear differences between two- and
    three-dimensional flows with the three-dimensional simulation
    showing much more small scale features.  This is consistent with
    the presence of relatively more power at larger wavenumber for the
    three-dimensional case compared to the two-dimensional case, as
    seen in Figure~\ref{fig:turb}.}
\end{figure}

Figure~\ref{fig:turb} also shows that there is relatively more power
in small scale (higher wavenumber, $k$) features for the
three-dimensional simulation compared to the two-dimensional
calculation.  This is made more explicit by looking at a colormap plot
of the enstrophy density $\eta = |\nabla\times\Ubt|^2/2$,
as is shown in Figure~\ref{fig:enstrophy}, where the left (right)
panel shows the two-dimensional (three-dimensional) simulation at
$t=0.02$~s.  The plot for the two-dimensional simulation is for the
wide domain, but only half of the domain is shown to keep the same
scale for both plots.  For the three-dimensional simulation, the plot
shows a slice through the center of the domain.  The two-dimensional
simulation plot appears to be dominated by moderate-sized vortices
throughout the domain, while in three dimensions, we see structure on
a much wider range of scales.  This is similar to the results seen in
comparisons of two- and three-dimensional simulations of
novae~\citep{kercek,kercek2}, although our two-dimensional results do
not show as severe of a dominance of vortices as reported there.  

The panels of Figure~\ref{fig:enstrophy} also show that in
two-dimensional flow the convective motions penetrate deeper into the
underlying neutron star than in the three-dimensional case.  This can
have implications for the amount of metal-rich material that can be
dredged up into the atmosphere, potentially polluting the photosphere
and adjusting the opacity.  We leave these details for a future paper.

\begin{figure}
  \centering
  \plotone{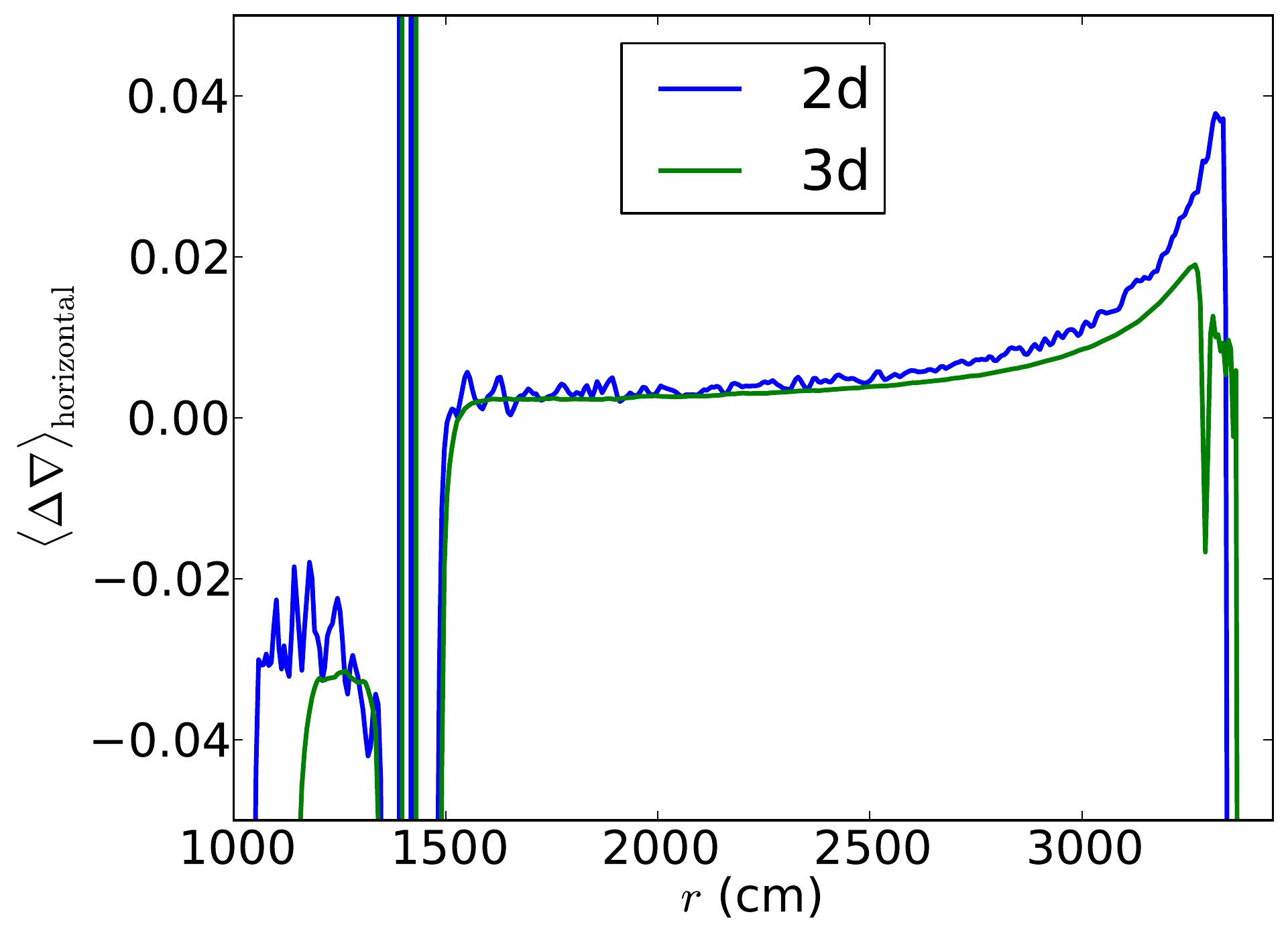}
  \caption{\label{fig:ad_excess} Horizontal average of the adiabatic
    excess, $\Delta\nabla$, at $t=0.02$~s for both two- and
    three-dimensional simulations.  The plot focuses near the
    convective region and shows the less extended overshoot region for
    the three-dimensional case, as seen in Figure~\ref{fig:enstrophy}.
    Furthermore, the upper convective boundary is quite different
    between the two simulations, with the two-dimensional model
    showing stronger superadiabaticity.}
\end{figure}

Accurate analysis of the convection during a thermonuclear runaway is
challenging.  Most convective analysis in the literature are focused
on stellar convection, which reaches a steady-state.  In that case,
one can drop the time derivative in the energy equation and simply
compare the balance of energy fluxes.  A thermonuclear runaway is far
from steady state.  One can assume things are in a quasi-steady state
over a somewhat short timescale and perform a RANS-like averaging of
the energy balance, but it is not {\em a priori} clear the exact
duration of this averaging timescale.  Our data dumps are roughly once every eddy
turnover time, which would likely not give good enough statistics for
this approach.  Instead, we have, as in \citet{xrb,xrb2},
focused on the adiabatic excess, $\Delta\nabla$:
\begin{equation}\label{eq:ad_excess}
  \Delta\nabla = \nabla - \nabla_s;\quad \nabla\equiv \frac{d\ln
    T/dr}{d\ln P/dr},
\end{equation}
where the subscript $s$ indicates the profile along an adiabat.
Figure~\ref{fig:ad_excess} shows the horizontal average of
$\Delta\nabla$ as a function of radius for both two- and
three-dimensional simulations at $t=0.02$~s.  This view of the
convective region confirms that the extent of the convective overshoot
region is less in three dimensions than in two dimensions, as was seen
in the comparison of Figure~\ref{fig:enstrophy}; in this snapshot, the
average overshoot region in two dimensions is roughly 50\% larger than
that of the three-dimensional simulation.  Furthermore, the upper
boundary in the two simulations is a bit different.  The two
dimensional simulation has a stronger degree of superadiabaticity,
implying the thermal gradient is steeper than that of an adiabat.  The
three-dimensional simulation also appears to have, on average, a
convectively stabilizing gradient around $r=3300$~cm where the
adiabatic excess becomes negative before a small overshoot region
extends the convection to nearly the same distance as in the
two-dimensional case.


\section{Discussion and Conclusions}\label{Sec:Conclusion}

We described the first three-dimensional models of convective burning
in an XRB.  While the peak temperature and Mach number behave
qualitatively the same as our two-dimensional calculations, the
structure of the convective velocity field differs substantially, both
in the global appearance and in the turbulent statistics.  This is
illustrated well by the difference in appearance of the enstrophy
density, which in three dimensions shows the typical cascade to small
scales.  Since convective mixing is expected to distribute the
synthesized nuclei throughout the atmosphere, potentially bringing
some to the photosphere, modeling the convection accurately is
important.  Based on the differences seen between the two-dimensional
and three-dimensional flows, this suggests that three-dimensional
models should be the focus of our future simulation efforts.

The calculations presented here pave the way for a more detailed study
of convective burning in XRBs.  We plan to do a more thorough analysis
of the convective flow patterns in both two and three dimensions in
the future paper by including tracer particles in the flow.  The
tracers will help visualize the trajectory of the flow, and to build a
statistical analysis of the transport during a thermonuclear runaway.
The wide three-dimensional simulation used 2.8 M CPU-hours on the OLCF
titan system (running with 768 MPI tasks and 16 threads per task).
While \maestro\ can use AMR, in these simulations we would refine the
entire convective region, so the cost savings would be small.
Modifying the simulation to advect and store tracer particle
information should increase the computational cost by only a few
percent.

Our future calculations will push for increased realism of the
reaction network.  As detailed in \citet{xrb2}, the approximate
network used here reasonably captures the overall energy release,
but we plan to both improve the nuclear reaction
network with a more clever selection of isotopes for an approximate
network, and to investigate using larger networks whose integration
can be accelerated using highly parallel hardware accelerators, such
as GPUs or Intel Xeon Phi processors.

Thus far, we have only explored a single initial model, constructed
with a simple parameterization.  The real state of the accreted layer
can vary, and there are two potential changes worth exploring.  First,
we extend the isentropic layer all the way to the surface of the
model, but accretion would likely cause heating at the top of the
atmosphere, which could truncate the convection region before the
surface.  This may prevent the convective plumes from reaching the
steep density gradient at the very top of the atmosphere.  Second, our
base density is on the higher end of likely models.  We should explore
lower density models as well.  The burning in that case would not be
as vigorous, but our timestep should increase as the convective
velocity decreases, making these simulations feasible.  An initial
study of the initial model variations can be done in two dimensions, and then
selected models can be run in three dimensions as needed.

We also plan to push our calculations to larger scales.  For the near
future, however, these sort of calculations will be limited to
convection-in-a-box studies.  Capturing the range of length scales
necessary to follow a laterally propagating burning front, while
resolving the energy-generation region, is not currently possible.
The complementary approach to ours are the calculations by
\citet{cavecchi:2012}, which used wide-aspect ratio zones and did not
perform hydrodynamics in the vertical direction.  Ultimately these two
methods can inform one-another to build a picture of nucleosynthesis
and dynamics of the burning front in XRBs.

\acknowledgements%

Visualization was done with {\sf yt} \citep{yt}.  The power spectrum
calculation followed the ``Making a Turbulent Kinetic Energy Power
Spectrum'' recipe in the {\sf yt} Cookbook.  The git-hashes of the
codes used for the main three-dimensional simulation are \maestro: {\tt
  afb7a1479b2b}$\ldots$ and \boxlib: {\tt 3fcc394f2774}$\ldots$.

We thank Frank Timmes for making his equation of state publicly
available.  The work at Stony Brook was supported by DOE/Office of
Nuclear Physics grants No.~DE-FG02-06ER41448 and DE-FG02-87ER40317 to
Stony Brook.  Work at LANL was done under the auspices of the National
Nuclear Security Administration of the U.S. Department of Energy at
Los Alamos National Laboratory under Contract No. DE-AC52-06NA25396.
The work at LBNL was supported by the Applied Mathematics Program of
the DOE Office of Advance Scientific Computing Research under
U.S. Department of Energy under contract No.~DE-AC02-05CH11231.  This
research used resources of the National Energy Research Scientific
Computing Center, which is supported by the Office of Science of the
U.S. Department of Energy under Contract No. DE-AC02-05CH11231.  An
award of computer time was provided by the Innovative and Novel
Computational Impact on Theory and Experiment (INCITE) program.  This
research used resources of the Oak Ridge Leadership Computing Facility
at the Oak Ridge National Laboratory, which is supported by the Office
of Science of the U.S. Department of Energy under Contract
No. DE-AC05-00OR22725.

\appendix

\section{Godunov Integration Details}\label{Sec:Godunov Integration Details}

Here we describe the details of the second-order Godunov integration
schemes used to predict face and time-centered quantities in various
steps of the algorithm.  In our overall algorithm, there are three
variations that share most of the same discretizations, with small
differences that will be described below.  In summary,
\begin{itemize}
\item {\bf Case 1:} {\em Construction of the advective velocities}
  
  In Steps 3 and 7 of the algorithm flowchart in \citet{MAESTRO:Multilevel}, we
  compute face and time-centered normal velocities (i.e., we only
  compute $u$ at $x$-faces, $v$ at $y$-faces, etc.), $\uadvstar$,
  given $\Ub^n$ and an associated source term, $\mathcal S_{\Ub}$.
  
\item {\bf Case 2:} {\em Construction of the final velocity edge states}

  In Step 11 of the \maestro\ flowchart, we compute face and
  time-centered velocities, $\Ub^{n+\myhalf}$, given $\Ub^n$ and an
  associated source term, $\mathcal S_{\Ub}$.  This case is different
  from {\bf Case 1} in that we leverage the availability of the
  projected velocity field, $\uadv$, during the characteristic tracing
  and upwinding steps.  Also, at each face we need to compute all
  components of velocity, rather than just the normal components.

\item {\bf Case 3:} {\em Construction of the scalar edge states}

  In Steps 4 and 8 of the \maestro\ flowchart, we compute face and
  time-centered scalars, $(\rho X_k, \rho h)^{n+\myhalf,(l)}$, given
  $(\rho X_k, \rho h)^n$ and the associated source terms,
  $\mathcal{S}_{\rho X_k}, \mathcal{S}_{\rho h}$.  This is different
  from {\bf Case 2} in that the evolution equations for the scalars
  are in conservative form rather than advective form.  
\end{itemize}

Our Godunov integration strategy is based on the piecewise parabolic
method (PPM) of \citet{ppm}.  We modify this algorithm to account for
the fact that (i) our underlying velocity field is spatially-varying,
(ii) we require a multidimensionally unsplit discretization, (iii) we
have governing equations in both advective {\it and} conservative form,
\begin{equation}
\frac{\partial q}{\partial t} = -\Ub\cdot\nabla q + \mathcal S_q; \quad q = u,v, ~\text{or}~ w,
\end{equation}
\begin{equation}
\frac{\partial q}{\partial t} = -\nabla\cdot(\Ub q) + \mathcal S_q; \quad q = \rho X_k ~\text{or}~ \rho h.
\end{equation}
Which form is used for the scalars, $(\rho X_k)$ and $(\rho h)$, is
determined at runtime based on how we chose to bring these states to
the interface.  \maestro\ offers several possibilities, e.g., bringing
$(\rho X_k)$ to the interface as a single quantity, bring $\rho'$ and
$X_k$ to the interface separately, or bringing $\rho$ and $X_k$ to the
interface separately.  Even more variation is allowed for $(\rho h)$,
where we can use $T$ instead of $h$ for the interface prediction.
We document both forms of interface reconstruction here.  The full
list of possible states is provided in the \maestro\ User's Guide.
For the present simulations, we predict $\rho$ and $X_k$ separately
for form $(\rho X_k)$ on interfaces, and $T$ is predicted and converted
to $h$ on the interface via the equation of state.

For all cases, the idea is to use the original PPM algorithm is to predict
preliminary face and time-centered states, $q^{\rm 1D}$, using a
one-dimensional advection equation for each direction $d$,
\begin{equation}
\frac{\partial q}{\partial t} = -U_d\frac{\partial q}{\partial x_d},
\end{equation}
and then use these states in a multidimensionally unsplit discretization of the full equations of motion
based on the ideas in \citet{ppmunsplit,saltzman1994} to compute updated face and time-centered states.
We now provide the details of our method.


\subsection{Case 1}
Here we compute face and time-centered estimates of the normal velocity.
We begin by computing preliminary face and time-centered estimates
of {\it all} velocity components at every face.
Here, $q$ will represent an arbitrary component of velocity ($u,v$, or $w$).
The following developments are for the $x$-direction and we omit the $j,k$
subscripts for ease of exposition; the equations for the $y$ and $z$-directions are analogous.
The first step is to construct a parabolic profile, $q(x)$, in each zone following the methodology
discussed in \citet{ppm}.  
\begin{equation}
  q(x) = q_{i,-} + \xi(x) (q_{i,+} - q_{i,-} + q_{6,i} (1 - \xi(x)))
\end{equation}
where $q_{i,-}$ and $q_{i,+}$ are the limited edge values of the parabola
and $q_{6,i} = 6q_{i}-3(q_{i,-}+q_{i,+})$ is related to the curvature.  The quantity $\xi(x)$ converts
$x$ into the fraction of the zone from the left edge.


The parabolic profiles are then integrated along characteristics to get the average value swept over the high
and low faces over the time step.  By defining
\begin{equation}
\sigma = |u_{i}|\frac{\Delta t}{\Delta x},\label{eq:case 1 sigma}
\end{equation}
we obtain:
\begin{eqnarray}
(q_L^{\rm 1D})_{i+\myhalf} &=&
\begin{cases}
q_{i,+} - \frac{\sigma}{2}\left[(q_{i,+}-q_{i,-})-\left(1-\frac{2}{3}\sigma\right)q_{6,i}\right],
& u_{i} > 0 \\
q_{i}, & \rm{otherwise}
\end{cases},\label{eq:case 1 ql}\\
(q_R^{\rm 1D})_{i-\myhalf} &=&
\begin{cases}
q_{i,-} + \frac{\sigma}{2}\left[(q_{i,+}-q_{i,-})+\left(1-\frac{2}{3}\sigma\right)q_{6,i}\right],
& u_{i} < 0 \\
q_{i}, & \rm{otherwise}
\end{cases}.\label{eq:case 1 qr}
\end{eqnarray}
Then, for the normal velocity components, we solve a Riemann problem to obtain
the preliminary state at the face,
\begin{equation}
u_{i+\myhalf}^{\rm 1D} =
\mathcal R\left((u_L^{\rm 1D})_{i+\myhalf},(u_R^{\rm 1D})_{i+\myhalf}\right) =
\begin{cases}
0, & (u_L^{\rm 1D})_{i+\myhalf} \le 0 ~~ {\rm AND} ~~ (u_R^{\rm 1D})_{i+\myhalf} \ge 0 \\
(u_L^{\rm 1D})_{i+\myhalf}, & (u_L^{\rm 1D})_{i+\myhalf} + (u_R^{\rm 1D})_{i+\myhalf} > 0 \\
(u_R^{\rm 1D})_{i+\myhalf}, & (u_L^{\rm 1D})_{i+\myhalf} + (u_R^{\rm 1D})_{i+\myhalf} < 0
\end{cases}.\label{eq:case 1 Riemann}
\end{equation}
Next, we obtain the preliminary face and time-centered transverse velocities 
using upwinding,
\begin{equation}
v_{i+\myhalf}^{1D} =
\mathcal U\left((v_L^{\rm 1D})_{i+\myhalf},(v_R^{\rm 1D})_{i+\myhalf},u_{i+\myhalf}^{\rm 1D}\right) =
\begin{cases}
(v_L)_{i+\myhalf}^{\rm 1D}, & u_{i+\myhalf}^{\rm 1D} > 0 \\
(v_R)_{i+\myhalf}^{\rm 1D}, & u_{i+\myhalf}^{\rm 1D} < 0
\end{cases}.
\end{equation}
We now have $u^{\rm 1D}$ and $v^{\rm 1D}$ at each face.
The next step is compute updated face and time-centered normal velocities by
accounting for the transverse derivative and source terms we have ignored so far.
In two dimensions, first compute $(u_L)_{i+\myhalf,j}$ and $(u_R)_{i+\myhalf,j}$ using, e.g.,
\begin{equation}
(u_L)_{i+\myhalf,j} = (u_L^{\rm 1D})_{i+\myhalf,j}
- \frac{\Delta t}{2}\left(\frac{v_{i,j+\myhalf}^{\rm 1D}+v_{i,j-\myhalf}^{\rm 1D}}{2}\right)\left(\frac{u^{\rm 1D}_{i,j+\myhalf}-u^{\rm 1D}_{i,j-\myhalf}}{\Delta y}\right)
+ \frac{\Delta t}{2}(\mathcal S_u)_{\ib},\label{eq:2D trans}
\end{equation}
followed by a Riemann solver to obtain the final face and time-centered state,
\begin{equation}
u_{i+\myhalf,j}^{\rm ADV,\star} = \mathcal R\left((u_L)_{i+\myhalf,j},(u_R)_{i+\myhalf,j}\right).
\end{equation}
In three dimensions, instead of Equation (\ref{eq:2D trans}) we use
\begin{eqnarray}
(u_L)_{i+\myhalf,j,k} &=& (u_L^{\rm 1D} )_{i+\myhalf,j,k}
- \frac{\Delta t}{2}\left(\frac{v_{i,j+\myhalf,k}^{\rm 1D}+v_{i,j-\myhalf,k}^{\rm 1D}}{2}\right)\left(\frac{u^{y|z}_{i,j+\myhalf,k}-u^{y|z}_{i,j-\myhalf,k}}{\Delta y}\right)\nonumber\\
&&\hspace{0.9in}- \frac{\Delta t}{2}\left(\frac{w_{i,j,k+\myhalf}^{\rm 1D}+v_{i,j,k-\myhalf}^{\rm 1D}}{2}\right)\left(\frac{u^{z|y}_{i,j,k+\myhalf}-u^{z|y}_{i,j,k-\myhalf}}{\Delta z}\right) 
+ \frac{\Delta t}{2}(\mathcal S_u)_{\ib},\nonumber\\
\end{eqnarray}
where to account for transverse corner coupling, we compute the intermediate states
as in \citet{ppmunsplit,saltzman1994}.  For example,
\begin{equation}
(u_L^{y|z})_{i,j+\myhalf,k} = (u_L)_{i,j+\myhalf,k} - \frac{\Delta t}{3}\left(\frac{w_{i,j,k+\myhalf}^{\rm 1D}+w_{i,j,k-\myhalf}^{\rm 1D}}{2}\right)\left(\frac{u_{i,j,k+\myhalf}^{\rm 1D}-u_{i,j,k-\myhalf}^{\rm 1D}}{\Delta z}\right),
\end{equation}
\begin{equation}
u_{i,j+\myhalf,k}^{y|z} = \mathcal U\left((u_L^{y|z})_{i,j+\myhalf,k},(u_R^{y|z})_{i,j+\myhalf,k},v_{i,j+\myhalf,k}^{\rm 1D}\right).
\end{equation}

\subsection{Case 2}
Here we compute face and time-centered estimates of each component of velocity at every face.
The details are the same as {\bf Case 1} up until Equation (\ref{eq:case 1 sigma}).
Now we can use $\uadv$ for characteristic tracing and upwinding whenever possible.
Specifically, we define
\begin{equation}
\sigma_\pm = \left|u^{\rm ADV}_{i\pm\myhalf}\right|\frac{\Delta t}{\Delta x},
\end{equation}
and compute
\begin{eqnarray}
(q_L^{\rm 1D})_{i+\myhalf} &=&
\begin{cases}
q_{i,+} - \frac{\sigma_+}{2}\left[(q_{i,+}-q_{i,-})-\left(1-\frac{2}{3}\sigma_+\right)q_{6,i}\right],
& u^{\rm ADV}_{i+\myhalf} > 0 \\
q_{i}, & \rm{otherwise}
\end{cases},\\
(q_R^{\rm 1D})_{i-\myhalf} &=&
\begin{cases}
q_{i,-} + \frac{\sigma_-}{2}\left[(q_{i,+}-q_{i,-})+\left(1-\frac{2}{3}\sigma_-\right)q_{6,i}\right],
& u^{\rm ADV}_{i-\myhalf} < 0 \\
q_{i}, & \rm{otherwise}
\end{cases},
\end{eqnarray}
\begin{equation}
q^{\rm 1D}_{i+\myhalf} = \mathcal U\left((q_L^{\rm 1D})_{i+\myhalf},(q_R^{\rm 1D})_{i+\myhalf},u^{\rm ADV}_{i+\myhalf}\right).\label{eq:case 2 upwind}
\end{equation}
We now have $q^{\rm 1D}$ at each face.
We now account for the transverse derivative terms to compute updated face and
time-centered states
$(q_L)_{i+\myhalf,j}$ and $(q_R)_{i+\myhalf,j}$ using, in two dimensions,
\begin{equation}
(q_L)_{i+\myhalf,j} = (q_L^{\rm 1D})_{i+\myhalf,j}
- \frac{\Delta t}{2}\left(\frac{v_{i,j+\myhalf}^{\rm ADV}+v_{i,j-\myhalf}^{\rm ADV}}{2}\right)\left(\frac{q^{\rm 1D}_{i,j+\myhalf}-q^{\rm 1D}_{i,j-\myhalf}}{\Delta y}\right)
+ \frac{\Delta t}{2}(\mathcal S_q)_{\ib},\label{eq:case 2 ql}
\end{equation}
followed by an upwinding step to obtain the final face and time-centered state,
\begin{equation}
q_{i+\myhalf,j}^{n+\myhalf} = \mathcal U\left((q_L)_{i+\myhalf,j},(q_R)_{i+\myhalf,j},u_{i+\myhalf,j}^{\rm ADV}\right).\label{eq:case 2 upwind final}
\end{equation}
In three dimensions we use
\begin{eqnarray}
(q_L)_{i+\myhalf,j,k} &=& (q_L^{\rm 1D} )_{i+\myhalf,j,k}
- \frac{\Delta t}{2}\left(\frac{v_{i,j+\myhalf,k}^{\rm ADV}+v_{i,j-\myhalf,k}^{\rm ADV}}{2}\right)\left(\frac{q^{y|z}_{i,j+\myhalf,k}-q^{y|z}_{i,j-\myhalf,k}}{\Delta y}\right)\nonumber\\
&&\hspace{0.9in}- \frac{\Delta t}{2}\left(\frac{w_{i,j,k+\myhalf}^{\rm ADV}+v_{i,j,k-\myhalf}^{\rm ADV}}{2}\right)\left(\frac{q^{z|y}_{i,j,k+\myhalf}-q^{z|y}_{i,j,k-\myhalf}}{\Delta z}\right) 
+ \frac{\Delta t}{2}(\mathcal S_q)_{\ib},\nonumber\\
\end{eqnarray}
where to account for transverse corner coupling, we compute the intermediate states
as in \citet{ppmunsplit,saltzman1994}.  For example,
\begin{equation}
(q_L^{y|z})_{i,j+\myhalf,k} = (q_L)_{i,j+\myhalf,k} - \frac{\Delta t}{3}\left(\frac{w_{i,j,k+\myhalf}^{\rm ADV}+w_{i,j,k-\myhalf}^{\rm ADV}}{2}\right)\left(\frac{q_{i,j,k+\myhalf}^{\rm 1D}-q_{i,j,k-\myhalf}^{\rm 1D}}{\Delta z}\right),
\end{equation}
\begin{equation}
q_{i,j+\myhalf,k}^{y|z} = \mathcal U\left((q_L^{y|z})_{i,j+\myhalf,k},(q_R^{y|z})_{i,j+\myhalf,k},v_{i,j+\myhalf,k}^{\rm ADV}\right).\label{eq:case 2 upwind trans}
\end{equation}

\subsection{Case 3}
Here we compute face and time-centered estimates of a $q$ that obeys a conservative equation.
The details are the same as {\bf Case 2} up through Equation (\ref{eq:case 2 ql}).  We
 also note that in {\bf Step 2A}, we use $\uadvpred$ instead of $\uadv$.  The difference 
between {\bf Case 2} is in the form of the corner coupling and transverse derivatives:
\begin{eqnarray}
(q_L)_{i+\myhalf,j} &=& (q_L)_{i+\myhalf,j}^{\rm 1D} - \frac{\Delta t}{2}\left(\frac{(q^{\rm 1D}v^{\rm ADV})_{i,j+\myhalf}-(q^{\rm 1D}v^{\rm ADV})_{i,j-\myhalf}}{\Delta y}\right) \nonumber\\
&&-\frac{\Delta t}{2}q_\ib\left(\frac{u_{i+\myhalf,j}^{\rm ADV} - u_{i-\myhalf,j}^{\rm ADV}}{\Delta x}\right)  + \frac{\Delta t}{2}(\mathcal S_q)_\ib.
\end{eqnarray}
We apply the same upwinding procedure in Equation (\ref{eq:case 2 upwind final}) to obtain the
final face and time-centered state, $q_{i+\myhalf,j}^{n+\myhalf}$.
In three dimensions we use
\begin{eqnarray}
(q_L)_{i+\myhalf,j,k} &=& (q_L^{\rm 1D} )_{i+\myhalf,j,k}
- \frac{\Delta t}{2}\left(\frac{(q^{y|z}v^{\rm ADV})_{i,j+\myhalf,k}-(q^{y|z}v^{\rm ADV})_{i,j-\myhalf,k}}{\Delta y}\right)\nonumber\\
&&\hspace{0.85in}- \frac{\Delta t}{2}\left(\frac{(q^{z|y}w^{\rm ADV})_{i,j,k+\myhalf}-(q^{z|y}w^{\rm ADV})_{i,j,k-\myhalf}}{\Delta z}\right) \nonumber\\
&&\hspace{0.85in}-\frac{\Delta t}{2}q_\ib\left(\frac{u_{i+\myhalf,j,k}^{\rm ADV} - u_{i-\myhalf,j,k}^{\rm ADV}}{\Delta x}\right) + \frac{\Delta t}{2}(\mathcal S_q)_\ib,
\end{eqnarray}
with
\begin{equation}
(q_L^{y|z})_{i,j+\myhalf,k} = (q_L)_{i,j+\myhalf,k} - \frac{\Delta t}{3}\left(\frac{(q^{\rm 1D}w^{\rm ADV})_{i,j,k+\myhalf}-(q^{\rm 1D}w^{\rm ADV})_{i,j,k-\myhalf}}{\Delta z}\right),
\end{equation}
along with the upwinding procedure in Equation (\ref{eq:case 2 upwind trans}).

\section{Two-Dimensional Tests}
Based on feedback from collaborators and the anonymous referee, we
have performed several two-dimensional XRB tests to see how either the
initial conditions or domain size can alter the effects of the
convection.

\begin{figure}
  \centering
  \includegraphics[width=0.48\textwidth]{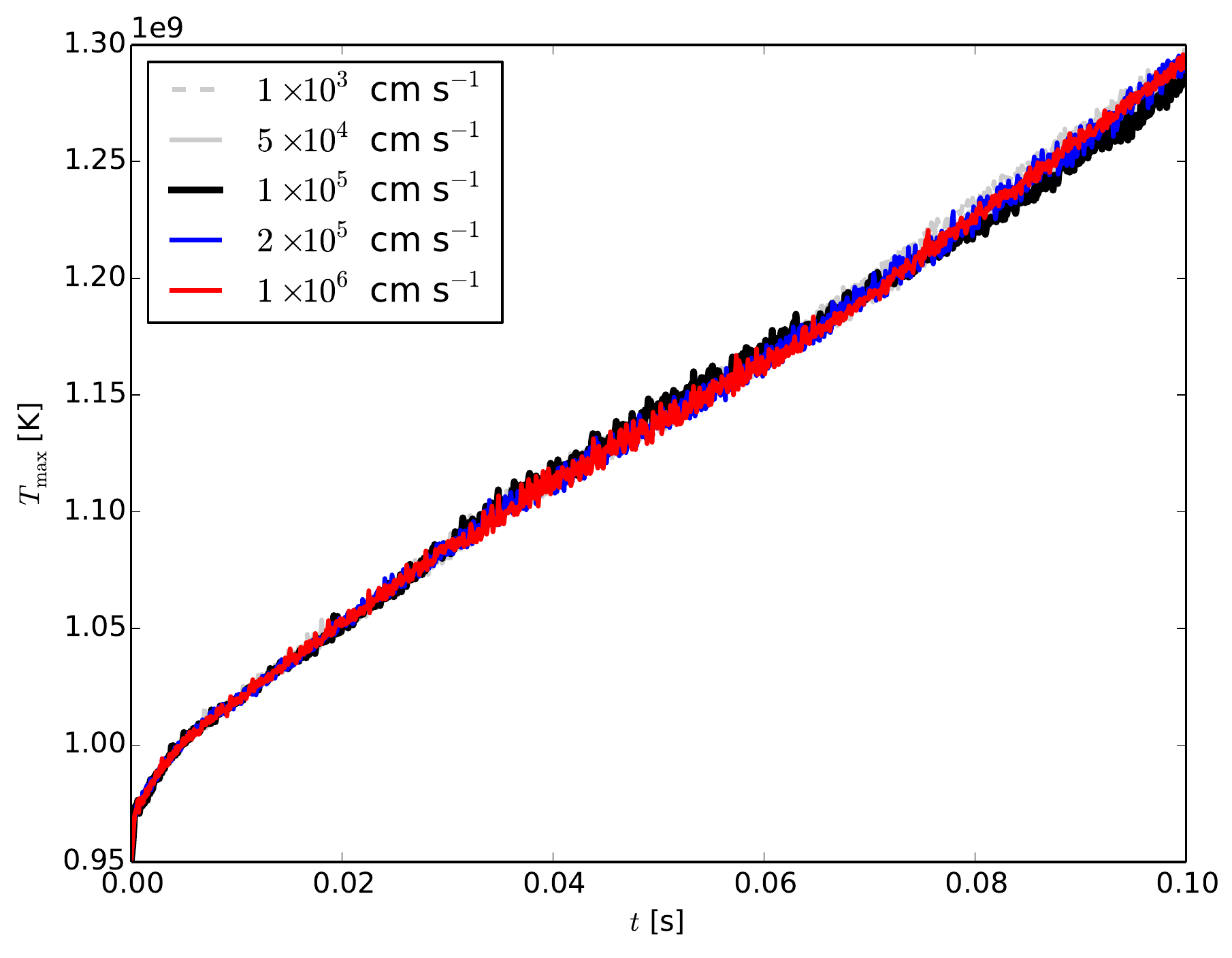}
  \includegraphics[width=0.48\textwidth]{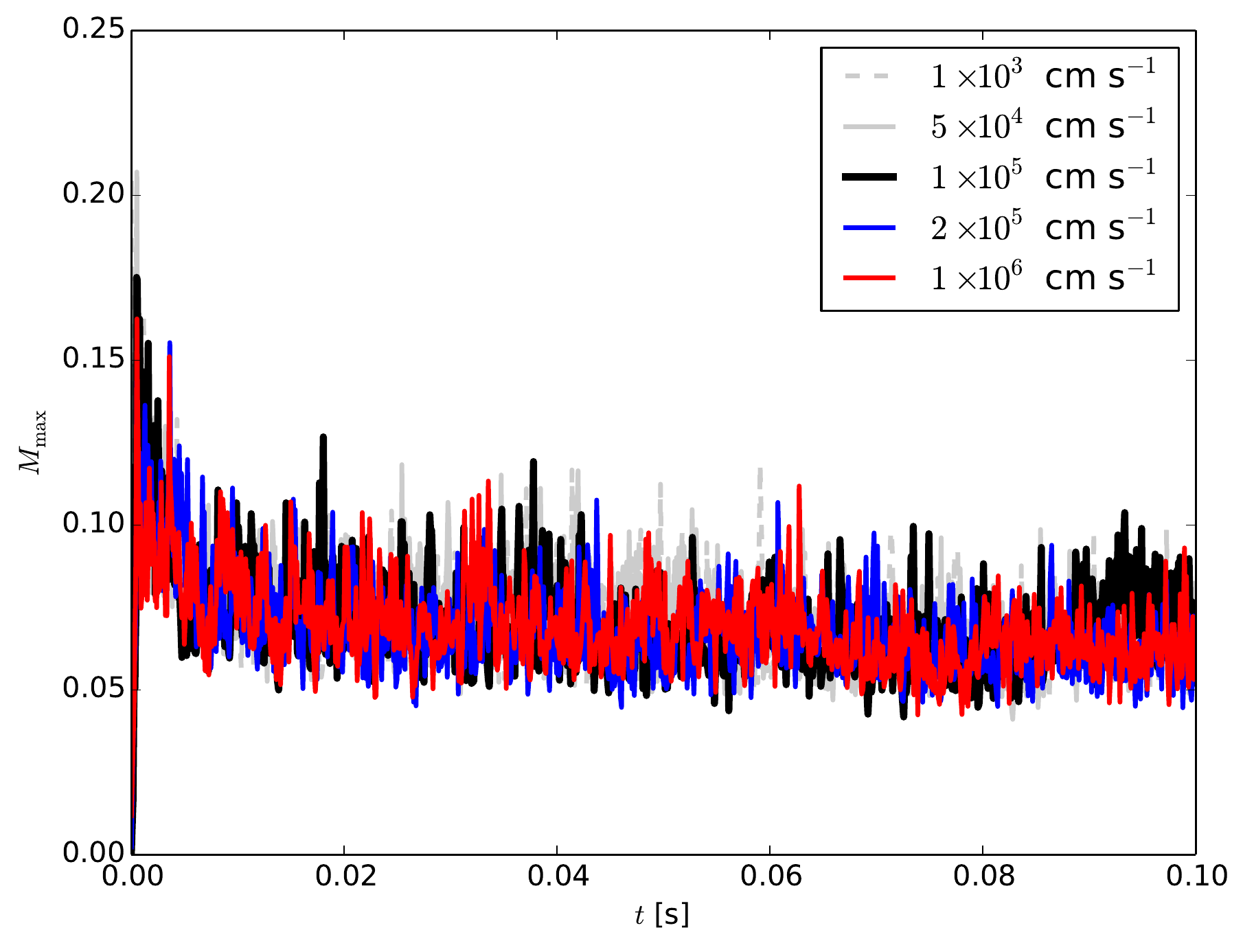}
  \caption{\label{fig:velpert_lines} Evolution of the peak temperature
    (left) and peak Mach number (right) as a function of time for five
    simulations that differ only in the strength of the initial
    velocity pertubation, which is given as the line labels.  All of
    the simulations track one another well, giving credence to the
    fact that the developed convection does not depend strongly on the
    initial pertubation strength.  Every 500$^\mathrm{th}$ point is
    plotted to minimize the image size.}
\end{figure}

\begin{figure}
  \centering
  \plotone{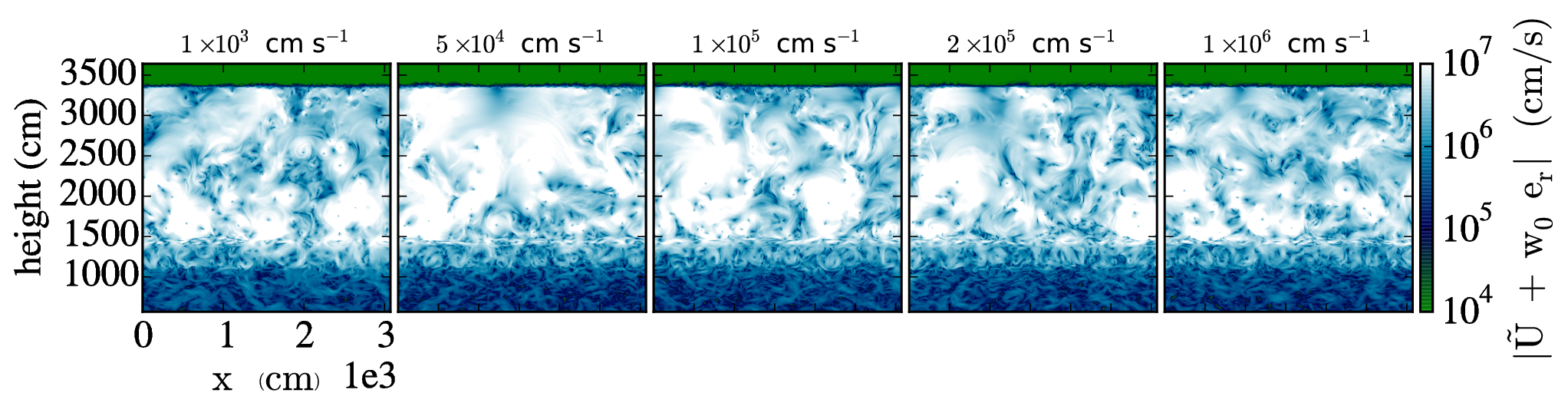}
  \caption{\label{fig:velpert_comp} Comparison of the magnitude
    of velocity field for the five simulations shown in Figure
    \ref{fig:velpert_lines} at $t=0.01$\ s; the velocity label along
    the top of each image gives the corresponding strength of the
    initial velocity perturbations.  The extent of the convective
    region as well as the rough magnitudes of the flow field are quite
    similar amongst the different runs.}
\end{figure}

The first test we performed was to alter the strength of the initial
velocity perturbations that act as a seed to the convection.  Even
though we expect the convection to ``forget'' how it was initiated
during the $\sim200$ convective turnover times we simulate, this is an
important check.  The default velocity perturbations in the
simulations of the main paper, as well as those of \cite{xrb2}, were
$1\times10^5$ cm s$^{-1}$.  Figure \ref{fig:velpert_lines} shows both
the peak temperature and peak Mach number as a function of time for
five simulations, each with different initial velocity perturbations
as shown in the labels; here we only plot every 500 steps to cut down
on the image size.  All simulations track each other very nicely, and
thus the development of the convection does not depend strongly on the
strength of the initial perturbations.  Figure \ref{fig:velpert_comp}
shows colormaps of the magnitude of velocity for each of the five
runs.  They likewise compare well with one another.

\begin{figure}
  \centering
  \plotone{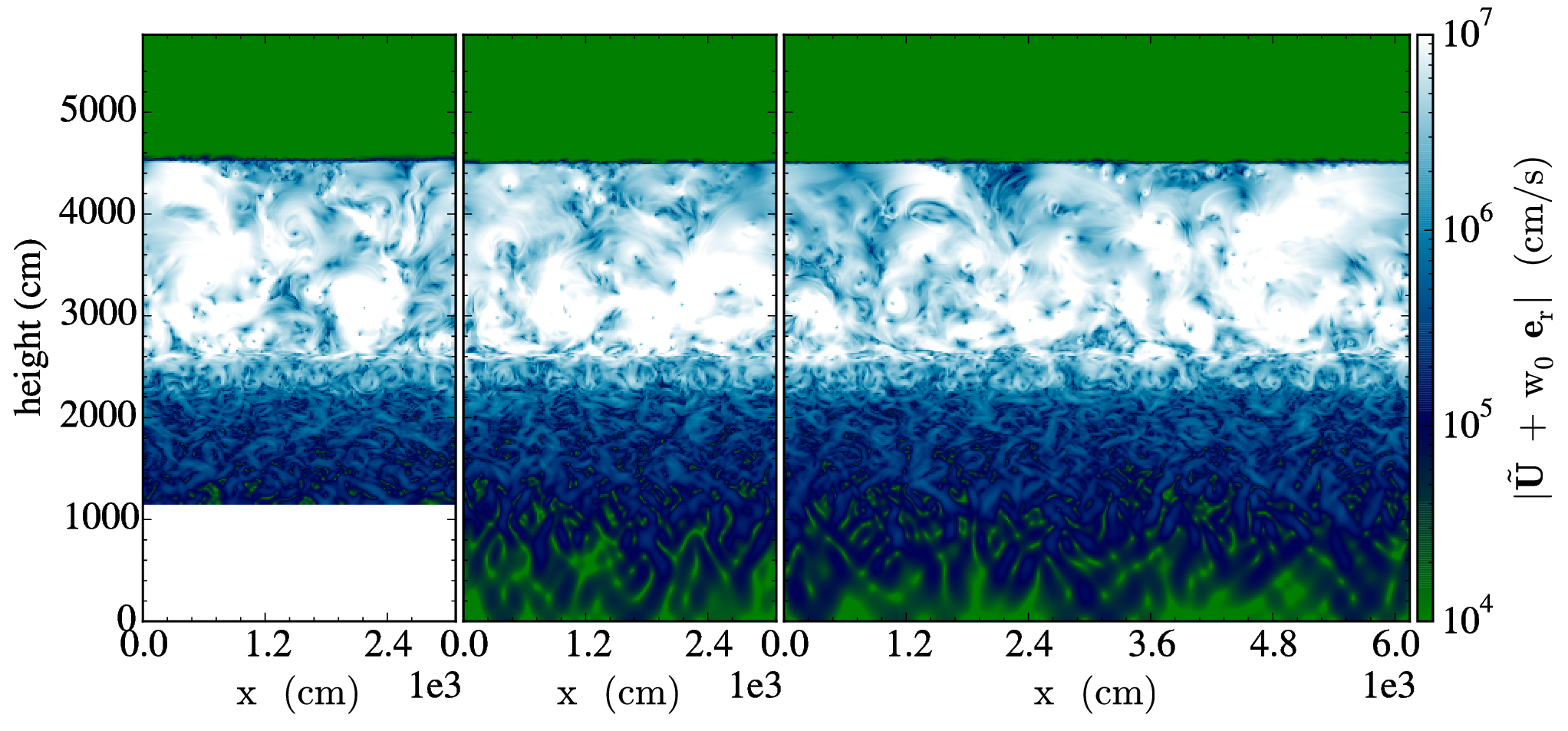}
  \caption{\label{fig:tall_comp} Comparison of the magnitude of
    velocity field for three simulations of varying size at
    $t=0.01$\ s.  The simulation on the left has the default size
    (3072 cm $\times$ 4608 cm), the middle simulation has the bottom
    of the domain extended to the size (3072 cm $\times$ 5760 cm), and
    the right simulation has the lateral extent of the domain extended
    to the size (6144 cm $\times$ 5760 cm).  The general intensities
    of the velocity field and the extent of the convective region are
    not affected by the default domain size we have chosen.}
\end{figure}

The second test was designed to see if the domain size choice we had
made for our default runs, including those of the three-dimensional
simulation, were affecting the evolution of the convective flow
pattern.  To this end, we performed two additional simulations where
we extended the default two-dimensional domain (3072 cm $\times$ 4608
cm) in both the vertical direction (to 3072 cm $\times$ 5760) and
again in the horizontal direction (to 6144 cm $\times$ 5760 cm).  The
results are shown in Figure \ref{fig:tall_comp} where we plot the
magnitude of velocity for each of the three runs at $t=0.01$\ s; the
default domain is on the left and has been shifted up by the 1152 cm
difference, the middle frame shows the tall domain, and the right
shows the tall and wide domain.  These plots show that the shape and
strength of the flow in the convective region are only weakly affected
by the domain size.  We note that had we chosen a domain width less
than the scale-height of the convection, this would have constrained
the eddy size quite strongly.


\clearpage



\clearpage

\clearpage

\clearpage

\clearpage

\clearpage

\clearpage

\clearpage

\clearpage

\clearpage

\end{document}